\RequirePackage{fix-cm}
\documentclass[twocolumn]{svjour3}          % twocolumn
\smartqed  % flush right qed marks, e.g. at end of proof
\usepackage{graphicx}
\begin{document}

\title{Raman Studies of Anisotropic Magnetic Excitations in Fluctuating Nematic 
Striped La$_{2-x}$Sr$_x$CuO$_4$ and the Comparison to Uniform Nd$_{2-x}$Ce$_x$CuO$_4$
}

\titlerunning{Raman studied of striped p- and uniform n-cuprates}        % if too long for running head

\author{S. Sugai \and Y. Takayanagi \and N. Hayamizu \and Y. Sone \and N. Nakagawa \and T. Muroi   %etc.
}


\institute{S. Sugai \at
              Department of Physics, Art and Science, Petroleum Institute, P.O. Box 2533, Abu Dhabi, UAE \\
              Tel.: +971-26075257\\
              \email{ssugai@pi.ac.ae}           %  \\
%              \emph{Present address:} of F. Author  %  if needed
           \\
              Department of Physics, Faculty of Science, Nagoya University, Furo-cho, Chikusa-ku, Nagoya 464-8602, Japan           \and
           Y. Takayanagi \at
              Department of Physics, Faculty of Science, Nagoya University, Furo-cho, Chikusa-ku, Nagoya 464-8602, Japan           \and
           N. Hayamizu \at
              Department of Physics, Faculty of Science, Nagoya University, Furo-cho, Chikusa-ku, Nagoya 464-8602, Japan           \and
           Y. Sone \at
              Department of Physics, Faculty of Science, Nagoya University, Furo-cho, Chikusa-ku, Nagoya 464-8602, Japan           \and
           N. Nakagawa \at
              Department of Physics, Faculty of Science, Nagoya University, Furo-cho, Chikusa-ku, Nagoya 464-8602, Japan           \and
           T. Muroi \at
              Department of Physics, Faculty of Science, Nagoya University, Furo-cho, Chikusa-ku, Nagoya 464-8602, Japan
}

\date{Received: date / Accepted: date}

\maketitle

\begin{abstract}
The mechanism of the high temperature hole-doped superconductivity was 
investigated by Raman scattering.  
The Raman selection rule is unique, so that anisotropic magnetic excitations 
in a fluctuating spin-charge stripe can be detected as if it is static.  
We use different Raman selection rules for two kinds of magnetic Raman 
scattering processes, two-magnon scattering and high-energy electronic 
scattering.  
In order to confirm the difference, the Raman spectra of striped 
La$_{2-x}$Sr$_x$CuO$_4$ (LSCO) and non-striped Nd$_{2-x}$Ce$_x$CuO$_4$ (NCCO) were 
compared.  
The main results in LSCO are (1) magnetic excitations are presented by individual 
energy dispersions for the $k||$ stripe and the $k\perp$ stripe, (2) the charge transfer 
is allowed only in the direction perpendicular to the stripe.  
The direction is the same as the Burgers vector of an edge dislocation.  
Hence we assume that a charge moves together with the edge dislocation of the charge 
stripe.  
The superconducting coherence length is close to the inter-charge stripe distance 
at $x<0.2$.  
Therefore we propose a model that superconducting pairs are formed in the edge 
dislocations.  
The binding energy is related to the stripe formation energy.  
\end{abstract}

\section{Introduction}
\label{intro}
The spin-charge stripe structure \cite{Yoshizawa,Birgeneau,Bianconi,Saini,Tranquada,Yamada,Matsuda,Tranquada2004,Matsuda2} in the hole-doped high temperature superconducting 
cuprate has been intensively investigated, because it has a strong possibility to 
solve the superconducting mechanism\cite{Zaanen,Kivelson,Zaanen2,Vojta}.  
In the previous conference (Superstripes 2011) we presented the direct observation 
of individual $k||$ and $k\perp$ stripe magnetic excitations in a fluctuating spin-charge 
stripe state utilizing high-energy Raman scattering \cite{Sugai2012}.  
The advantage of Raman scattering is the unique selection rule which is determined by two 
Cartesian vectors parallel to the electric fields of incident and scattered light.  
If we chooses the electric field of incident light to one of the possible stripe 
direction and the electric field of scattered light to the other possible stripe 
direction, the observed spectra do not depend on the stripe direction, because 
Raman scattering is symmetric for the exchange of incident and scattered light.  
Using this technique we can observe the fluctuating stripe, as if it is static.  
Magnetic scattering arises from two mechanisms, two-magnon scattering and high-energy 
electronic scattering.  
The two mechanisms have different symmetries \cite{Sugai2012}.  
Using the difference we disclosed the following results.  
(1) The magnetic excitations in the stripe state of LSCO are presented by the 
dispersions calculated by Seibold and Lorenzana \cite{Seibold,Seibold2}.  
The energy of the $k||$ stripe dispersion decreases together with the decrease 
of the high-energy intensity in the magnetic susceptibility.  
The separation of the dispersion curve in the $k\perp$ stripe is caused by the 
Brillouin zone folding due to the extended magnetic unit cell.  
(2) the electronic scattering spectra show only $k\perp$ stripe excitations, 
indicating that the carrier hopping is restricted to the perpendicular 
direction to the stripe.  

In order to confirm the above analysis, we 
compared the Raman spectra of striped LSCO and non-striped electron-doped 
NCCO \cite{YamadaPRL2003}.  
The experimental results certified the above analysis in LSCO.  
The restriction of the charge transfer direction is reminiscent of the Burgers vector 
of an edge dislocation.  
The edge dislocation easily slide in the Burgers vector direction which is 
perpendicular to an inserted half layer, giving ductility in metal.  
The carrier density dependent coherence length which gives the pair size 
is close to the inter-charge stripe distance at $x<0.2$.  
It indicates that the charge moves together with the edge 
dislocation and the superconducting pairs are formed in the edge dislocations.  

High energy magnetic Raman scattering has been reported in 
hole doped superconductors \cite{Blumberg,Naeini,Nachumi,Sugai,Machtoub,Tassini,Muschler} 
and electron-doped superconductors \cite{Sugai2,Tomeno,Onose}.  
Many of them reported only the B$_{\rm 1g}$ spectra, because the B$_{\rm 2g}$ 
spectra are two-magnon scattering inactive.  
The B$_{\rm 2g}$ high-energy electronic Raman spectra are very sensitive to the crystal 
surface condition.  
Only fresh cleaved surface gives the key structure, a hump from 1100 to 3100 cm$^{-1}$ 
in LSCO. 
The hump is scarcely reported except for a sign in the report by Machtoub \cite{Machtoub}.  
The spectra in NCCO is sensitive to the oxygen reduction.  
We optimized the reduction condition to shorten the superconducting transition range 
less than 2 degree in the resistivity curve.

\section{High-energy Raman spectra in striped LSCO and non-striped NCCO}
\label{sec:1}
\begin{figure*}
\includegraphics[width=\textwidth]{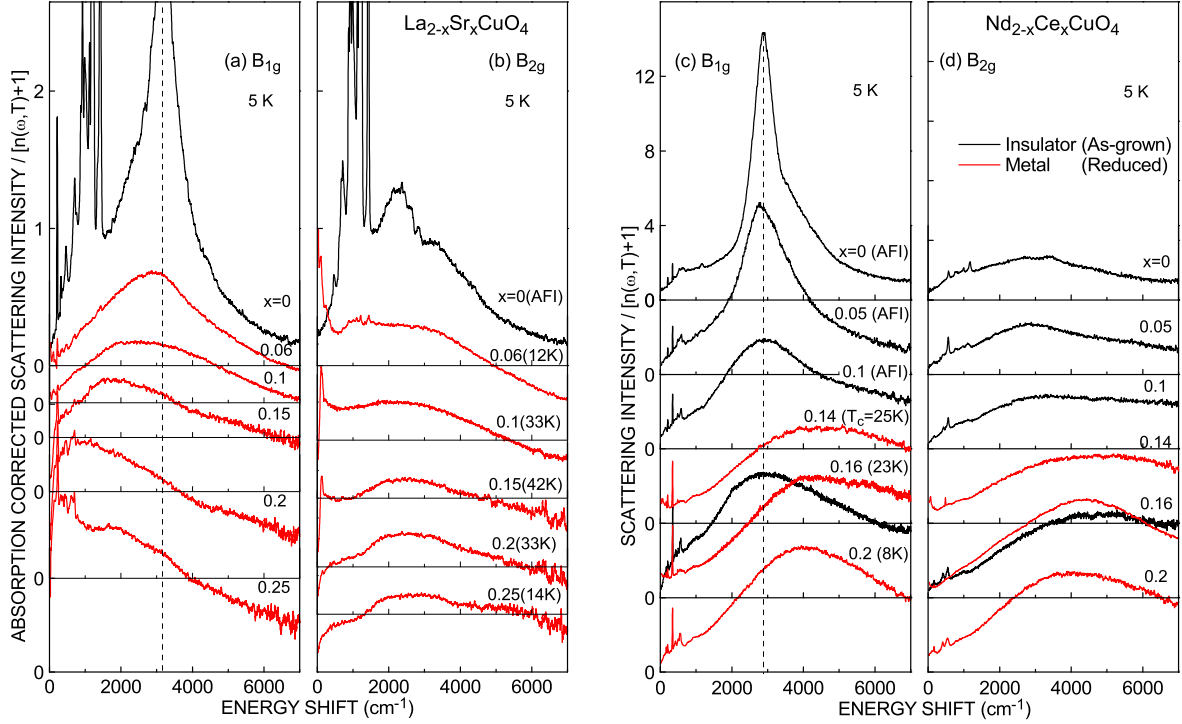}
\caption{(color online) 
$B_{\rm 1g}$ and $B_{\rm 2g}$ Raman spectra in LSCO and NCCO.  
The spectra in the insulating phase is shown by 
the black curve and in the metallic phase by the red curve.  
The black spectra in NCCO were measured in as-grown crystals and the red spectra in 
reduced crystals.  
The two-magnon peak energies are shown by the dashed lines. 
The sharp peaks below 1400 cm$^{-1}$ in La$_2$CuO$_4$ are two-phonon peaks.
}
\label{fig:1}
\end{figure*}

Figure 1 shows the comparison of Raman spectra in LSCO and NCCO.  
The spectra in insulator are shown by black curves and those in metal by red curves.  
The high-energy $B_{\rm 1g}$ spectra show magnetic excitations caused by the 
two-magnon scattering process and the electronic scattering process, while the 
$B_{\rm 2g}$ spectra present magnetic excitations caused by the electronic 
scattering process only.  
In LSCO the $B_{\rm 1g}$ spectra represent $k||$ and $k\perp$ stripe excitations and the 
$B_{\rm 2g}$ spectra only $k\perp$ stripe excitations.  
In LSCO the $B_{\rm 1g}$ magnetic scattering peak whose energy decreases with 
increasing $x$ is caused by the $k||$ dispersion.  
The hump from 1100 cm$^{-1}$ to 3100 cm$^{-1}$ in the $B_{\rm 1g}$ and $B_{\rm 2g}$ 
spectra is caused by the $k\perp$ dispersion.  

Each of $B_{\rm 1g}$ and $B_{\rm 2g}$ spectra in NCCO are almost the same as in LSCO 
at $x=0$, because the magnetic scattering mechanism is only two-magnon scattering.  
However, the spectra in the metallic phase are very different.  
The black curve in Fig. 1(b) shows the scattering in as-grown insulating 
crystals and the red curve in oxygen reduced metallic crystals.  
The $B_{\rm 1g}$ peak energy in the as-grown crystal does not change even if Ce 
concentration increases to $x=0.16$ where the reduced sample is a superconductor.  
It is noted that the $B_{\rm 2g}$ spectra change as Ce concentration increases 
regardless of as-grown or reduced.  
When the crystal becomes metal, the $B_{\rm 1g}$ spectra abruptly shift to high 
energy and change into nearly the same form as the $B_{\rm 2g}$ spectra.  
The isotropic charge transfer in the uniform spin lattice gives the same spectra in 
$B_{\rm 1g}$ and $B_{\rm 2g}$ symmetries.  
The peak energy can be interpreted by the theoretical model that an electron hops 
in a uniform antiferromagnetic spin lattice \cite{Manousakis}.  
A hopping electron overturns the site spin, so that the motion of an electron can 
be approximated by the motion in the linear confining potential increasing with the path 
length.  The transition energies between the discrete levels are 4800, 8200,... 
cm$^{-1}$ using $J=0.3t$ and $t=0.4 eV$.  
The lowest transition energy coincides with the energy of the broad Raman peak in 
the metallic phase at $x=0.14$.  

Thus the Raman spectra in the metallic phase are very different by the existence or 
absence of the spin-charge stripe structure.  
The $B_{\rm 1g}$ two-magnon peak shifts to low energy in LSCO, as carrier density 
increases.  
On the other hand in NCCO it keeps the constant energy in the insulating phase even 
if Ce concentration increases and two-magnon peak disappears in the metallic phase.  
The $B_{\rm 2g}$ hump from 1100 to 3100 cm$^{-1}$ in LSCO which is assigned to 
the dispersion segment in the $k\perp$ stripe does not appear in NCCO.  
These experimental results certify our assignment that the $B_{\rm 2g}$ spectra in 
LSCO present the $k\perp$ stripe excitations.

\section{Charge transfer united with the edge dislocation of the stripe in LSCO}
The surprising result of the charge transfer only in the perpendicular direction to 
the stripe is reminiscent of an edge dislocation in metal.  
The edge dislocation easily slides perpendicularly to the inserted layer.  
Applying the properties of the edge dislocation to the stripe state, the experimental 
results are interpreted that carriers in the charge stripe are localized except for 
the edge and the conductivity is induced by the hopping of charges united with the 
edge dislocation.  

\begin{figure}
\includegraphics[width=0.5\textwidth]{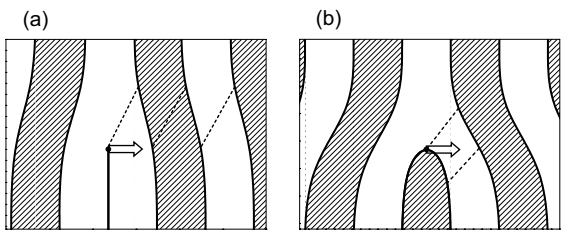}
\caption{(a) The single edge dislocation and (b) the lopped edge dislocation.  
The hatched area and the white area denote spin stripes with opposite phases and 
the boundary is the charge stripe.  The arrow denotes the Burgers vector which is 
the sliding direction of the edge dislocation.  
The dashed line segments show the movement of the charge stripes at the sliding 
of the edge dislocation.
}
\label{fig:3}
\end{figure}

Figure 2(a) shows the single edge dislocation and Fig. 2(b) the looped edge 
dislocation \cite{Zaanen}.  
The spin alignments at both sides of the charge stripe have opposite 
phases \cite{Tranquada}, so that the looped edge dislocation 
has lower energy \cite{Zaanen}.  
The edge dislocation moves to right by displacing small parts of charge stripes at 
and near the edge indicated by the dashed segments.  
The density of the edge dislocation is assumed to increase with increasing the 
carrier density.  
As for the superconductivity in the stripe phase, the Bosonization of charge and 
spin dynamics in a one-dimensional conductor (Tomonaga-Luttinger liquid) was 
used \cite{Zaanen,Zaanen2}.  
However, the separation of the spin and charge degree of freedom has not been observed 
and the charge hopping in the looped edge dislocation is not a simple one-dimensional 
hopping.  
Therefore we suppose that the superconducting pairs are formed in the moving 
carriers at the edge dislocation.  
This model is supported by the carrier density dependent coherence length in the 
following.  

\section{Superconducting pairs formed in the edge dislocation}
\begin{figure*}
\includegraphics[width=0.75\textwidth]{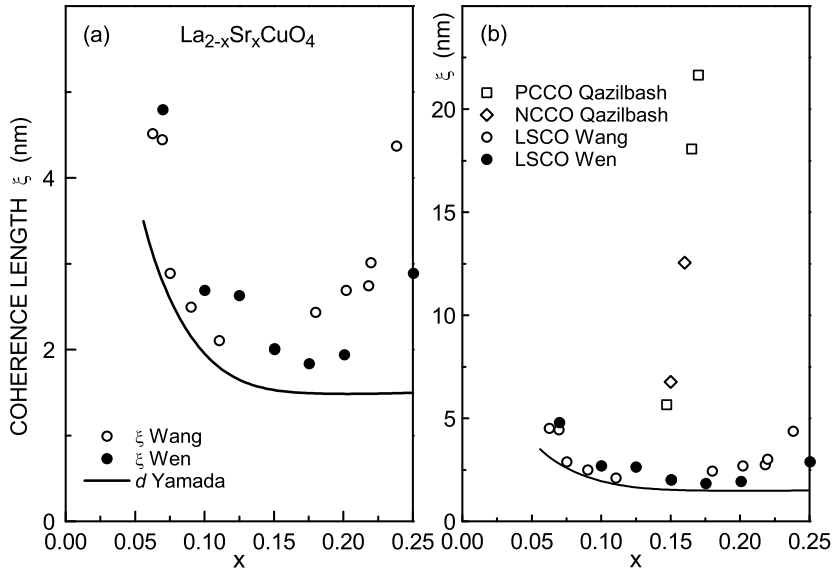}
\caption{(a) The coherence length $\xi$ \cite{Wang,Wen} and the inter-charge stripe distance 
$d$ in LSCO \cite{Yamada} and (b) the $\xi$ in NCCO, PCCO \cite{Qazilbash}, and 
LSCO \cite{Wang,Wen} and the $d$ in LSCO \cite{Yamada}.
}
\label{fig:3}
\end{figure*}

Figure 3(a) shows the carrier density dependence of the Ginzburg-Landau coherence 
length $\xi=\sqrt{\Phi_0/2\pi H_{c2}}$ \cite{Wang,Wen} and the distance $d$ \cite{Yamada} 
between charge stripes in LSCO, where $\Phi_0$ is the fluxoid quantum.  
The $d$ obtained from neutron scattering decreases from the insulator-metal transition 
point to $x=1/8$ and then keeps constant at $x>1/8$.  
The $\xi$ is close to $d$ from the insulator-metal transition point to $x\approx 0.2$ 
and then increases as $x$ increases.  
The increase of $\xi$ at $x>0.2$ may be caused by the shortening of stripes by the 
increased edge dislocation density.  

Figure 3(b) shows the coherence length $\xi$ 
of hole-doped LSCO \cite{Wang,Wen} and electron-doped NCCO and PCCO \cite{Qazilbash}.  
The coherence lengths of NCCO and PCCO are sited on a single curve.  
The $\xi$ of NCCO and PCCO is much longer than that of LSCO.  
The coherence length $\xi=22$ nm at $x=0.17$ 
approaches $\xi=38$ nm of Nb.  
The stronger the binding energy is, the shorter the  coherence length is.  
The clear difference between the striped and the non-striped cuprates indicates that 
the superconducting mechanism is related to the stripe in the hole-doped cuprate and 
different in the electron-doped cuprate.  

Thus the proximity between $\xi$ and $d$ in LSCO indicates that the pairing is 
formed in the moving edge dislocation.  
The coherence length of LSCO at $x<0.2$ is only twice the inter-hole distance 
on the assumption that holes are uniformly distributed.  
Therefore the superconducting state is in the crossover regime between BCS (Bardeen-
Cooper-Schrieffer) and BEC(Bose-Einstein condensation).  
It is like a bipolaron \cite{Alexandrov}, but the binding energy is related to the 
stripe formation energy rather than the electron-phonon interaction energy and the 
charge transfer direction is restricted to one direction for each pair.  

In summary two kinds of magnetic Raman scattering mechanisms in the stripe state 
is confirmed by comparing the spectra in LSCO and NCCO.  
The magnetic excitations in LSCO are expressed by the individual energy dispersions 
in the $k||$ stripe and the $k\perp$ stripe.  
The charge transfer is restricted in the direction perpendicular to the stripe.  
This direction is the same as the Burgers vector of the edge dislocation.  
Hence the charge transfer is assumed to be united with the edge dislocation.  
We propose a model that the superconducting pairs are formed in the edge dislocations, 
because the coherence length is close to the distance between the charge stripes.


\begin{thebibliography}{}
\bibitem{Yoshizawa}H. Yoshizawa, S. Mitsuda, H. Kitazawa, and K. Katsumata, J. Phys. 
Soc. Jpn., {\bf 57}, 3686 (1988). 

\bibitem{Birgeneau}R. J. Birgeneau, Y. Endoh, K. Kakurai, Y. Hidaka, T. Murakami, 
M. A. Kastner, T. R. Thurston, G. Shirane, and K. Yamada, Phys. Rev. B {\bf 39}, 
2868 (1989).

\bibitem{Bianconi}A. Bianconi, N. L. Saini, A. Lanzara, M. Missori, T. Rossetti, 
H. Oyanagi, H. Yamaguchi, K. Oka, and T. Ito, 
Phys. Rev. Lett. {\bf 76}, 3412 (1996).

\bibitem{Saini}N. L. Saini, H. Oyanagi, T. Ito, V. Scagnoli, M. Filippi, S. Agrestini, 
G. Campi, K. Oka, and A. Bianconi, Eur. Phys. J. B {\bf 36}, 75 (2003).

\bibitem{Tranquada}J. M. TranquadaCB. J. SternllebCJ. D. AxeCY. Nakamura, 
and S. Uchida, Nature {\bf 375}, 561 (1995).

\bibitem{Yamada}K. Yamada, C. H. Lee, K. Kurahashi, J. Wada, S. Wakimoto, S. Ueki, 
H. Kimura, Y. Endoh, S. Hosoya, G. Shirane, R. J. Birgeneau, M. Greven, M. A. Kastner, 
and Y. J. Kim, Phys. Rev. B {\bf 57}, 6165 (1998).

\bibitem{Matsuda}M. Matsuda, M. Fujita and K. Yamada, R. J. Birgeneau, Y. Endoh, 
and G. Shirane, Phys. Rev. B, {\bf 65}, 134515 (2002).

\bibitem{Tranquada2004} J. M. Tranquada, H. Woo, T. G. Perring, H. Goka, G. D. Gu, 
G. Xu, M. Fujita, and K. Yamada, Nature {\bf 429}, 534 (2004).

\bibitem{Matsuda2}M. Matsuda, M. Fujita, S. Wakimoto, J. A. Fernandez-Baca, 
J. M. Tranquada, and K. Yamada, Phys. Rev. Lett. {\bf 101}, 197001 (2008).

\bibitem{Zaanen}J. Zaanen, O. Y. Osman, H. V. Kruis, Z. Nussinov,and J. Tworzydlo, 
 Philos. Mag. B {\bf 81}, 
1485 (2001).

\bibitem{Kivelson}S. A. Kivelson, I. P. Bindloss, E. Fradkin, V. Oganesyan, 
J. M. Tranquada, A. Kapitulnik, and C. Howald, Rev. Mod. Phys., {\bf 75}, 
1201 (2003).

\bibitem{Zaanen2}J. Zaanen, Z. Nussinov, and S. I. Mukhin, Ann. Phys. ({\it NY}) 
{\bf 310}, 181 (2004). 

\bibitem{Vojta}M. Vojta, Adv. Phys., {\bf 58}, 699 (2009).

\bibitem{Sugai2012}S. Sugai, Y. Tamai, Y. Takayanagi, N. Hayamizu, and T. Muroi, 
J. Supercond. Nov. Magn. {\bf 25}, 1393 (2012).  

\bibitem{Seibold}G. Seibold, and J. Lorenzana, Phys. Rev. B {\bf 73}, 
144515 (2006).

\bibitem{Seibold2}G. Seibold, and J. Lorenzana, 
Phys. Rev. B {\bf 80}, 012509 (2009).

\bibitem{YamadaPRL2003}K. Yamada, K. Kurahashi, T. Uefuji, M. Fujita, S. Park, 
S.-H. Lee, and Y. Endoh, Phys. Rev. Lett. {\bf 90}, 137004 (2003).

\bibitem{Blumberg}G. Blumberg, M. Kang, M. V. Klein, K. Kadowaki, C. Kendzior, 
Science, {\bf 278}, 1427 (1997).

\bibitem{Naeini}J. G. Naeini,  J. C. Irwin,T. Sasagawa, Y. Togawa, and K. Kishio, 
Canadian J. Phys. {\bf 78}, 483 (2000). 

\bibitem{Nachumi}B. Nachumi, C. Kendziora, N. Ichikawa, Y. Nakamura, S. Uchida, 
Phys. Rev. B {\bf 65}, 092504 (2002). 

\bibitem{Sugai}S. Sugai, H. Suzuki, Y. Takayanagi, T. Hosokawa, 
N. Hayamizu, Phys. Rev. B {\bf 68}, 184504 (2003).

\bibitem{Machtoub}L. H. Machtoub, B. Keimer, and K. Yamada, Phys. Rev. 
Lett. {\bf 94}, 107009 (2005).

\bibitem{Tassini}L. Tassini, W. Prestel, A. Erb, M. Lambacher, R. Hackl, 
Phys. Rev. B {\bf 78}, 020511(R) (2008).

\bibitem{Muschler}B. Muschler, W. Prestel, L. Tassini, R. Hackl, M. Lambacher, 
A. Erb, S. Komiya, Y. Ando, D. C. Peets, W. N. Hardy, R. Liang, and D. A. Bonn, 
Eur. Phys. J. Special Topics {\bf 188}, 131 (2010).

\bibitem{Sugai2}S. Sugai, Y. Hidaka, Phys. Rev. B {\bf 44}, 809 (1991). 

\bibitem{Tomeno}I. Tomeno,M. Yoshida, K. Ikeda, K. Tai, K. Takamuku, N. Koshizuka, 
S. Tanaka, K. Oka, and H. Unoki, Phys. Rev. B {\bf 43}, 3009 (1991).

\bibitem{Onose}Y. Onose, Y. Taguchi, K. Ishizaka, Y. Tokura, Phys. Rev. B 
{\bf 69}, 024504 (2004).

\bibitem{Manousakis}E. Manousakis, Phys. Rev. B {\bf 75}, 035106 (2007).

\bibitem{Wang}Y. Wang, and H.-H. Wen, Europhys. Lett. {\bf 81}, 57007 (2008).

\bibitem{Wen}H. H. Wen, H. P. Yang, S. L. Li, X. H. Zeng, A. A. Soukiassian, 
W. D. Si, and X. X. Xi, Europhys. Lett. {\bf 64}, 790 (2003).

\bibitem{Qazilbash}M. M. Qazilbash, A. Koitzsch, B. S. Dennis, A. Gozar, 
Hamza Balci, C. A. Kendziora, R. L. Greene, and G. Blumberg, Phys. Rev. B {\bf 72}, 
214510 (2005).

\bibitem{Alexandrov}A. S. Alexandrov, J. Ranninger, and S. Robaszkiewicz, Phys. Rev. B 
{\bf 33}, 4526 (1986).  

\end{thebibliography}
\end{document}